# SIMULATION OF METAL-OXIDE MELT INTERACTION IN VIEW OF KINETICS OF CHEMICAL REACTIONS IN THE INTERPHASE BOUNDARY

## M. Zinigrad

*Ariel University, Israel*

Thermodynamics analysis of oxidation-reduction reactions between metal melt and slag (1) provides answers to certain practical issues such as the path of specific chemical reactions, final (equilibrium) phase composition, and the elements that are reduced and oxidized at given physical parameters. Although considerable, this is obviously not enough to analyze real technological systems, because the required equilibrium cannot be normally achieved despite high temperatures of welding and metallurgical processes. Hence, a dynamic problem has to be resolved here, which is calculating phase composition as a function of time. This cannot be achieved without knowing the rates of element concentration changes in each and every phase, and also technological parameters of the process. Relevant studies are detailed in(2-19).

This paper analyzes special kinetic features of physical and chemical processes in the metal and oxide melt interface.

We will start with a relatively simple case, which is deriving a kinetic equation for a monomolecular reaction. Let us assume that the following hypothetical process proceeds in the interphase boundary:

$$B^{(1)} \rightarrow F^{(2)}, \qquad (1)$$

where $B$ and $F$ are the 1st and 2nd phase components, respectively.

This reaction can be described as follows (see Fig. 1).
Rates of all subsequent stages in the steady mode are equal:

$$V = V_d^B = V_r = V_d^F, \qquad (2)$$

where $V$ – is heterogeneous reaction rate (1);

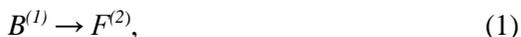

$V_r$, $V^B_d$ и $V^F_d$ – are the rates of chemical reaction at the phase boundary, diffusion rates of reagent $B$ in solution 1 and reagent $F$ in solution 2, respectively.

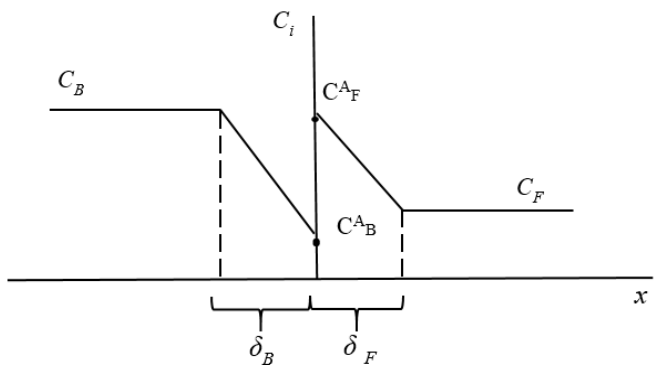

**Fig. 1**. Scheme of the hypothetical process $B^{(1)} \to F^{(2)}$. $C_B$, $C_F$ – are component concentrations in the solution; $C^A_B$, $C^A_F$ –– are component concentrations in the interface; $\delta_B$, $\delta_F$ – is the thickness of the diffused layer indicative of special features of the molecular diffusion of components.

As known from formal kinetics, the heterogeneous reaction rate in the activation mode is described by the following kinetic formula:

$$V = K_r C_B^A, \qquad (3)$$

where $K_r$ – is the constant of chemical reaction rate.

At the same time, as follows from the 1st Fick's law, rate equation for the diffusion mode can be as follows:

$$V_d^B = \frac{D_B}{\delta_B}. \qquad (4)$$

In general, the kinetic equation for heterogeneous monomolecular reaction will be as follows:

$$V = \frac{K_r \cdot \dfrac{D_B}{\delta_B}}{K_r + \dfrac{D_B}{\delta_B}} \cdot C_B, \qquad (5)$$

where $D_B$ is diffusion coefficient.

Such an approach is obviously limited as it can not be applied to assess the rate of multimolecular reactions. Transition to multimolecular processes certainly makes this problem more complicated.

**Kinetics of multimolecular reaction**

Now let us consider a heterogeneous reaction in the metal-oxide melt interface:

$$[Mn] + (FeO) \rightarrow (MnO) + [Fe]. \qquad (6)$$

$$K_{Mn} = \frac{(MnO)^{eq} \cdot [Fe]^{eq}}{[Mn]^{eq} \cdot (FeO)^{eq}}, \qquad (7)$$

where   parentheses–is mass % of components in the oxide phase;
square brackets – is mass % of components in the metal phase.
Scheme of the process on can see in Fig. 2.

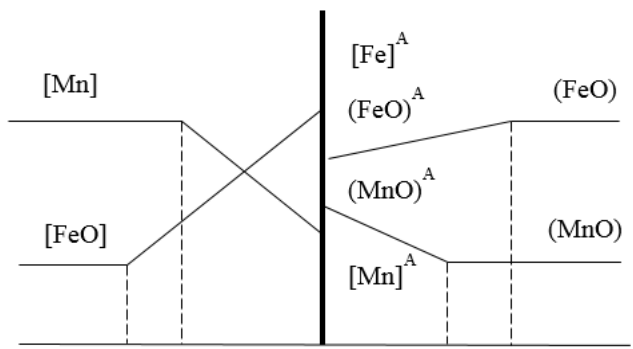

**Fig. 2.** Scheme of the heterogeneous process.

At constant temperatures, reaction rate depends on reagent concentrations:
$$V = f([Mn], (MnO), [Fe], (FeO)).$$

Concentration change rates of all reagents in the steady mode are equal:
$$V = V_{Mn} = V_{FeO} = V_{Fe} = V_{MnO}. \qquad (8)$$

Assuming that iron activity in the metal melt as a basic element ($a_{Fe}$) equals 1, we obtain a simplified formula for constant equilibrium:

$$K_{Mn} = \frac{(MnO)^{eq}}{[Mn]^{eq} \cdot (FeO)^{eq}}. \qquad (9)$$

Further, we assume that the process runs in the diffusion mode. In such a case, interface concentrations are close to equilibrium with the diffusion rate of each component depending on diffusion parameters and the difference between bulk and interface concentrations (i.e., in the state of equilibrium):

$$V_i = \frac{D_i}{\delta_i}(C_i - C_i^A). \qquad (10)$$

More specifically, we have the following rate equations for various reagents for the process selected (6), with $\frac{D_i}{\delta_i}$ designated as $K_i^d$ (diffusion rate constant) in the case of diffusion control:

$$V_{Mn} = k_{Mn}^D ([Mn] - [Mn]^{eq}) = k_{Mn}^D [Mn]\left(1 - \frac{[Mn]^{eq}}{[Mn]}\right), \qquad (11)$$

$$V_{FeO} = k_{FeO}^D [(FeO) - (FeO)^{eq}] = k_{FeO}^D (FeO)\left[1 - \frac{(FeO)^{eq}}{(FeO)}\right], \qquad (12)$$

$$V_{MnO} = k_{MnO}^D [(MnO)^{eq} - (MnO)] = k_{MnO}^D (MnO)\left[\frac{(MnO)^{eq}}{(MnO)} - 1\right], \qquad (13)$$

$$V_j^{Lim} = k_j^a c_j. \qquad (14)$$

Allowing for (14), the equation for reagent equilibrium concentrations will be as follows:

$$[Mn]^{eq} = [Mn]\left(1 - \frac{V_{Mn}}{V_{Mn}^{\lim}}\right), \qquad (15)$$

$$[FeO]^{eq} = [FeO]\left(1 - \frac{V_{FeO}}{V_{FeO}^{\lim}}\right), \qquad (16)$$

$$[MnO]^{eq} = [MnO]\left(1 - \frac{V_{MnO}}{V_{MnO}^{\lim}}\right). \qquad (17)$$

Now, we insert (15), (16) and (17) into the equilibrium constant formula (9):

$$K_{Mn} = \frac{(MnO)\left(1 + \dfrac{V_{MnO}}{V_{MnO}^{\lim}}\right)}{[Mn]\left(1 - \dfrac{V_{Mn}}{V_{Mn}^{\lim}}\right) \cdot (FeO)\left(1 - \dfrac{V_{FeO}}{V_{FeO}^{\lim}}\right)}. \qquad (18)$$

Therefore, the kinetic equation of the multimolecular reaction is as follows:

$$K_{Mn} = \frac{(MnO)\left(1 + \dfrac{V_{MnO}}{V_{MnO}^{\lim}}\right)}{[Mn]\left(1 - \dfrac{V_{Mn}}{V_{Mn}^{\lim}}\right) \cdot (FeO)\left(1 - \dfrac{V_{FeO}}{V_{FeO}^{\lim}}\right)}, \qquad (19)$$

Because the concentration change rates are the same for all reagents:

$$V = V_{MnO} = V_{Mn} = V_{Mn} = V_{FeO}.$$

For convenience, this example deals with equal stoichiometric factors for all reagents in the reaction. More particularly, for reaction:

$$4[Al] + 3\,(SiO_2) \rightarrow 3[Si] + 2\,(Al_2O_3). \qquad (20)$$

Kinetic equation will be as follows:

$$K_{Al} \frac{[Al]^4 \cdot (SiO_2)^3}{[Si]^3 \cdot (Al_2O_3)^2} = \frac{\left(1+\dfrac{3V}{V_{Si}^{\lim}}\right)^3 \left(1+\dfrac{2V}{V_{Al_2O_3}^{\lim}}\right)^2}{\left(1-\dfrac{4V}{V_{Al}^{\lim}}\right)^4 \left(1-\dfrac{3V}{V_{SiO_2}^{\lim}}\right)^3}. \quad (21)$$

**Simulation of Phase Interaction Kinetics in a Multicomponent System**

Let us consider a simple technological process (Fig. 3)

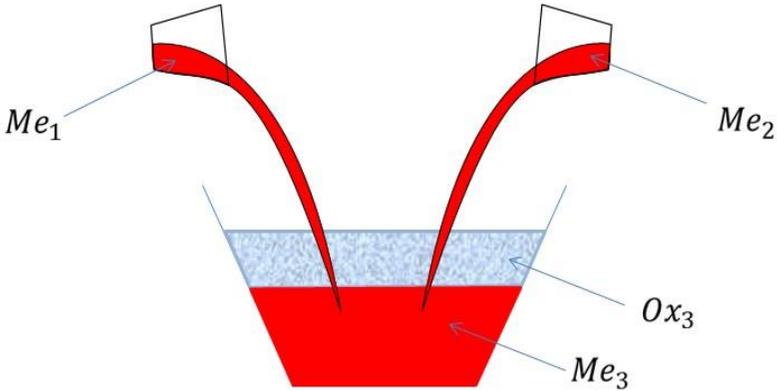

**Fig. 3.** Technological process chart.

Metal melt in vessel 3 is formed by concentrations of components in vessels 1 and 2 as well as chemical reaction with the oxide. Metal mass in vessel 3 within (*Δt*) time period can be easily calculated as follows:

$$m_{Me} = v_1 \Delta t + v_2 \Delta t + m_3, \quad (22)$$

where $m_{M3}$ is the initial mass, $v_1$ and $v_2$ are the rates of metal feed from vessels 1 and 2.

Thus, any element $X$ concentration, as achieved due to a simple mixing, disregarding a chemical reaction, can be determined as follows:

$$[E_i]_{mix}^{\Delta t} = \frac{[E_i]_1 \cdot v_1 \cdot \Delta t + [E_i]_2 \cdot v_2 \cdot \Delta t + [E_i]_3 \cdot m_3}{m}, \quad (23)$$

where $[E_i]_1$, $[E_i]_2$ and $[E_i]_3$ are the $i^{th}$ element concentrations in vessels 1, 2, and 3 at the start of the process.

Mass ($m_{E_i}^{\Delta t}$) and concentration ($[E_i]_r^{\Delta t}$) of the $i^{th}$ element formed due to chemical reactions between metal and oxide phases are determined as follows:

$$m_{E_i}^{\Delta t} = V_{E_i} \cdot \Delta t \cdot A, \quad (24)$$

$$[E_i]_r^{\Delta t} = \frac{100\% \cdot V_{E_i} \cdot \Delta t \cdot A}{m}, \quad (25)$$

where $V_{E_i}$ - is the $i^{th}$ element concentration change rate due to chemical reaction; $A$–is the interphase surface.

The total concentration change is determined as follows:

$$[E_i]^{\Delta t} = [E_i]_{mix}^{\Delta t} + [E_i]_r^{\Delta t}. \quad (26)$$

The equations, which would describe component concentration changes in both phases, are as follows:

$$[E_i]^{\Delta t} = \frac{[E_i]_1 \cdot v_1 \cdot \Delta t + [E_i]_2 \cdot v_2 \cdot \Delta t + [E_i]_3 \cdot m_{Me_3} + 100 \cdot V_{E_i} \cdot A \cdot \Delta t}{v_1 \cdot \Delta t + v_2 \cdot \Delta t + m_{Me_3}}, \quad (27)$$

$$(E_{in}O_m)^{\Delta t} = (E_{in}O_m) + \frac{100\% \cdot V_{E_i} \cdot \Delta t \cdot A}{m_{ox}^n}. \quad (28)$$

Equations (27) and (28) contain a most important parameter that has a significant effect on the phase formation. This is the rate of element transition through interphase boundary $V_{E_i}$.

The authors apply their own method for kinetic analysis of multimolecular and simultaneous reactions (19, 20, 8). The general equation to describe interaction of each of the metal melt components with the same reagent in the oxide melt will be as follows:

$$\frac{n}{m}[E_i] + (FeO) \rightarrow \frac{1}{m}(E_{in}O_m) + [Fe], \quad V_i \left[ \frac{molFeO}{cm^2 \cdot S} \right]. \quad (29)$$

The common reagent in (29) is FeO.

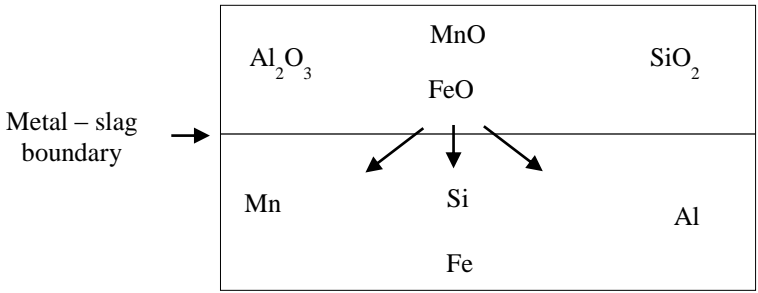

**Fig. 4.** Metal component interaction with the common reagent FeO.

According to (20), the theoretical basis of the method consists of two assumptions:
1) In a diffusion-controlled mode, concentration ratio in the interface for each reaction is close to the equilibrium value.
2) The reagent transition rate to or from the interface is proportional to the difference between concentrations in the bulk and in the metal-oxide melt interface.

Use of the approach proposed in (20) enabled the authors to derive the equations for the rates of transition of any number of elements through the metal-oxide melt interface ($V_{Ei}$) with mutual influence of all relevant reactions taken into account.

As follows from the above chart (Fig. 4), iron oxide flow is distributed among all components in the metal melt:

$$V_{FeO} = V_1 + V_2 + \cdots + V_i = \sum V_i = \sum \frac{m}{n} V_{Ei}. \qquad (30)$$

Iron oxide $(V_{FeO})$ and metal melt component $(V_{Ei})$ flows are formulated as follows (20):

$$V_{FeO} = \frac{\dfrac{(FeO)}{[Fe]} - x}{\dfrac{x}{V_{Fe}^{\lim}} + \dfrac{(FeO)}{[Fe] \cdot V_{FeO}^{\lim}}}. \qquad (31)$$

If stoichiometric factor $n = 1$:

$$V_{Ei} = \frac{x^m - K_{Ei}^m \dfrac{(E_{in}O_m)}{[E_i]}}{\dfrac{x^m}{V_{Ei}^{\lim}} + \dfrac{K_{Ei}^m (E_{in}O_m)}{[E_i] V_{E_{in}O_m}^{\lim}}}. \qquad (32)$$

If $n = 2$:

$$V_{E_i} = V_{E_i}^{lim}\left[1 + \emptyset \frac{V_{E_i}^{lim}}{4V_{Ei2Om}^{lim}} - \sqrt{\left(1 + \emptyset \frac{V_{E_i}^{lim}}{4V_{Ei2Om}^{lim}}\right)^2 - 1 + \varphi}\,\right] \quad (33)$$

Rates of all subsequent stages in the steady mode are equal:

$$V = V_d^B = V_r = V_d^F, \qquad (2)$$

where $V$ – is heterogeneous reaction rate (1);

$V_r$, $V^B{}_d$ è $V^F{}_d$ – are the rates of chemical reaction at the phase boundary, diffusion rates of reagent $B$ in solution 1 and reagent $F$ in solution 2, respectively.

$$\varphi = \frac{K_{Ei}^m (E_{i2}O_m)}{x^m [E_1]^2}.$$

281

By solving the equations (27), (28), (30), (31), (32) and (33), we can determine phase composition at any time and for any number of components.

In fact, this set of equations is a mathematical model of oxide-metal phase interaction for a selected technological process (Fig. 3).

Hence, phase composition equations are supposed to correlate with a specific technological process, since mass ratio and technological parameters have a significant effect on the nature of chemical reactions.

This approach was applied, for example, to develop mathematical models for steel treatment with slag in a ladle (21), building up technology (22), submerged arc welding (23), tungsten-containing waste treatment technology (24), and other metallurgical technologies (25-41).

29. V.Mazurovsky, M. Zinigrad, A. Zinigrad: 'Novel computer-aided method of welding materials design'. Computer Technology in Welding, AWS, USA 2001 11.
30. M. Zinigrad, V. Mazurovsky: 'Development of new welding materials on the base of mathematical modeling of metallurgical processes Part 1. Phase interaction analysis and development of the basic model'. In the book 'The optimization of composition, structure and properties of metals, oxides, composites, nano and amorphous materials'. Bi-National Russia-Israel Workshop, Ekaterinburg, Russia 2002 277-291.
31. M. Zinigrad, V. Mazurovsky: 'Development of new welding materials on the base of mathematical modeling of metallurgical processes. Part 2. Development of solution algorithm and software'. In the book 'The optimization of composition, structure and properties of metals, oxides, composites, nano and amorphous materials'. Bi-National Russia-Israel Workshop, Ekaterinburg, Russia 2002 292-303.
32. V. Mazurovsky, M. Zinigrad, A. Zinigrad: 'Mathematical Model of Weld Microstructure Formation'. Proceedings of the 12$^{th}$ International Conference 'Computer Technology in Welding and Manufacturing, TWI, Sydney, Australia 2002 12 79/1-79/9.
33. M. Zinigrad, V. Mazurovsky, A. Zinigrad: 'Mathematical modeling of phase interaction taking place during fusion welding processes'. Yazawa International Symposium, San Diego, USA 2003 667- 680.
34. V. Mazurovsky, M. Zinigrad, A. Zinigrad, L. Leontiev and V. Lisin: New approach to welding materials design'. In the book 'The optimization of composition, structure and properties of metals, oxides, composites, nano and amorphous materials'. Bi-National Israel-Russia Workshop, Jerusalem, Israel, 2003 144-154.
35. V. Mazurovsky, M. Zinigrad, A. Zinigrad, L. Leontiev, V. Lisin: 'The phenomenological model of non-equilibrium crystallization and strengthening-phase-formation processes in the weld'. In the book 'The optimization of composition, structure and properties of metals, oxides, composites, nano and amorphous materials'. Bi-National Israel-Russia Workshop, Jerusalem, Israel 2003 155-167.
36. V. Mazurovsky, M. Zinigrad, L. Leontev, V. Lisin: 'Physicochemical analysis and modeling of the primary crystallization processes of a metal during welding'. In the book 'The optimization of composition,